\documentclass[12pt,a4paper]{article}
\usepackage{amssymb}
\usepackage{amsmath}
\usepackage{amsfonts}
\usepackage{graphicx}
\usepackage{comment}
\usepackage[hyperindex=true,
          pdfstartview=FitH,
          bookmarksnumbered=true,
          bookmarksopen=true,
          citecolor=blue,
          linkcolor=blue,
          colorlinks=true,
          pdfborder=001,
          unicode]{hyperref}

\begin{document}
\title{Novel accelerating Einstein vacua and smooth inhomogeneous 
Riemannian manifolds}
\author{Kun Meng${}^{1}$\thanks{email: kunmeng@mail.nankai.edu.cn}, Wei Xu${}^{1}$\thanks{email: xuweifuture@mail.nankai.edu.cn} and Liu Zhao${}^{1,2}$
\thanks{email: lzhao@nankai.edu.cn}\\
${}^{1}$School of Physics, Nankai University, Tianjin 300071, China\\
${}^{2}$Kavli Institute for Theoretical Physics China, \\
CAS, Beijing 100190, China}
\date{}
\maketitle
\begin{abstract}
A novel class of Einstein vacua is presented, which possess non-vanishing cosmological 
constant and accelerating horizon with the topology of $S^{D-3}$ fibration over 
$S^{1}$. After Euclideanization, the solution describes
a conformally distorted $S^{D-1}$ fibration over $S^1$, which is smooth, compact and 
inhomogeneous, and can be regarded as analogue of Don Page's 
gravitational instanton.
\end{abstract}

\section{Introduction}

Despite its nearly a hundred-year-old age, Einstein gravity continues to be 
a source of inspirations and surprises. Besides the great success in describing physics 
in solar system, Einstein gravity also predicts existence of various black holes and even 
extended objects like black strings \cite{Horne:1991gn,Anderson:1995hu,
Mahapatra:1993gx} and black rings \cite{Emparan:2001p607,
Emparan:2001p762,Elvang:2004p3129,Emparan:2006p2455} in higher dimensions. 
Not all aspects 
of these nontrivial solutions have been fully understood. Even in the pure vacuum 
sector, Einstein gravity has been shown to possess an unexpected richness -- in addition 
to the well known maximally symmetric vacua (dS, AdS, Minkowski etc), Einstein
gravity also admits inhomogeneous vacua such as the anisotropic accelerating vacua 
\cite{Zhao:2011p693,Zhao:2011p1230,Xu:2011p1323} and 
the massless topological black hole vacua \cite{Mann:1996p968,
Vanzo:1997p1331,Birmingham:1998p1113}
and so on. In this article, we shall present a novel 
class of Einstein vacua which possess accelerating horizons of nontrivial topology. 
Concretely, the vacua we shall be discussing are accelerating vacua with horizons being 
conformally distorted sphere bundles over $S^{1}$. After Euclideanization, the whole 
vacuum metric becomes that of a conformally distorted sphere bundle over $S^{1}$, 
which corresponds to a smooth, compact and inhomogeneous Riemannian manifold.

Historically, the first known smooth, compact and inhomogeneous Riemannian manifold
of constant scalar curvature was the gravitational instanton discovered by Don Page
\cite{Page:1978p1271}. 
Page's gravitational instanton describes a nontrivial metric of $S^{2}$ bundle over 
$S^{2}$. Such metrics were later generalized to higher dimensions 
\cite{Hashimoto:2004p1291}, giving rise to 
constant curvature metrics of  nontrivial $S^{D-2}$ bundle over $S^{2}$. 
In differential geometry, sphere bundles over $S^{1}$ are more complicated than 
sphere bundles over $S^{2}$, because the latter are simply connected and are known 
to be Einstein manifolds, while the former are non-simply connected and are not 
Einstein manifolds. The present work shows that, although $S^{D-1}$ bundles 
over $S^{1}$ are in general not Einstein, there exists an Einstein metric in the 
conformal class of such manifolds.

\section{A 5D vacuum solution}

We begin our study by presenting a novel Einstein vacuum solution in five dimensions
(5D). We start from 5D not because 5D is of any particular importance for the 
construction, but because we began studying this subject with the aim of finding black 
rings with cosmological constant. Though we have not yet fulfilled our aim, the result 
presented here indeed bears some resemblance to black ring solutions, with the 
exception that instead of black ring horizon of topology $S^{2}\times S^{1}$, we now 
have a cosmological horizon of the same topology.

The metric of the novel vacuum solution is given as follows,
\begin{align}
  \mathrm{d} s^2=\frac{1}{(1+\alpha r\cos\theta)^2}\left[-h(r) \mathrm{d} t^2
  +\frac{\mathrm{d} r^2}{h(r)}+r^2 \mathrm{d} \theta^2
  +r^2\sin^2\theta \mathrm{d} \psi^2
  +(L+r\cos\theta)^2 \mathrm{d} \phi^2 \right], \label{5dm}
\end{align}
where
\begin{align}
  h(r)=1-\frac{\alpha r^2}{L}. \label{hr}
\end{align}
This metric is a generalized version of a 4-dimensional metric found in 
\cite{Astorino:2011p1101}. 
It is straightforward to show that this metric is an exact solution to the vacuum Einstein 
equation with a cosmological constant given by
\begin{align*}
\Lambda=\frac{6\alpha(1-\alpha L)}{L}. \label{lambda}
\end{align*}
For $L>0, \alpha>0$ and $\alpha L<1$, this is an inhomogeneous de Sitter spacetime
with no singularities. For $\alpha L>1$, the metric is still free of essential singularities,
but the cosmological constant becomes negative, and there exist apparent singularities in 
the conformal factor which indicates non-compactness of the spacetime in this case.  
In the rest of this article, we assume $\alpha L<1$, i.e. $\Lambda>0$.

The singularity free nature of the metric is best manifested by the calculation of curvature invariants. For instances, we have
\begin{align*}
&R_{\mu\nu\rho\sigma}R^{\mu\nu\rho\sigma}=
\frac{10\Lambda^{2}}{9},\\
&R_{\mu\nu}R^{\mu\nu}=\frac{20\Lambda^{2}}{9},\\
&R=\frac{10\Lambda}{3}.
\end{align*}

Physically, the metric describes an accelerating vacuum of Einstein gravity with the accelerating horizon taking a nontrivial topology. The proper acceleration $a^\mu=
u^\nu \nabla_\nu u^\mu$ for the static observers in the spacetime has  the norm
\[
a^\mu a_\mu=\left(\frac{\alpha}{L}\right)^2\left(L^2+
\frac{r^2-2rL\cos\theta+\alpha L r^2\cos^2\theta}{1-\frac{\alpha}{L}r^2}\right).
\]
This quantity has the finite value $\alpha^2$ at $r=0$ and blows up to infinity at
$r=\sqrt{\frac{L}{\alpha}}$. So the $r_H=\sqrt{\frac{L}{\alpha}}$ hyper surface represents 
an accelerating horizon. Notice that the coordinate $r$ is not the radial variable in polar 
coordinate system, $r=0$ corresponds not to the spacial origin but to a circle of radius 
$L$. Thus, unlike the usual de Sitter spacetime, the acceleration 
horizon is a topologically nontrivial manifold. 

\subsection{Horizon geometry} \label{hg}

To understand the nontrivial topology of the acceleration horizon, we now look at the 
metric on the horizon surface. We have
\begin{align*}
  \mathrm{d} s_H^2=\frac{1 }{(1+\alpha r_H\cos\theta)^2}\bigg\{
 r_H^2( \mathrm{d} \theta^2+\sin^2\theta \mathrm{d} \psi^2)
  +\left(L+r_H\cos\theta\right)^2 \mathrm{d} \phi^2 \bigg\}.
\end{align*}
Let us temporarily put the conformal factor aside. The 3D hyper surface with the metric 
\begin{align}
\mathrm{d}\tilde s^{2}= r_H^2( \mathrm{d} \theta^2+\sin^2\theta \mathrm{d} 
\psi^2)
  +\left(L+r_H\cos\theta\right)^2 \mathrm{d} \phi^2 \label{dst}
\end{align}
has a very nice geometric interpretation. Let us follow the treatment of \cite{Frolov:2006pu} of this geometry.  
Consider a global embedding of a 3D hypersurface
\begin{align}
  x^2+y^2+(\sqrt{z^2+w^2}-L)^2=r_H^2  \label{hs}
\end{align}
into 4D Euclidean space $\mathbb{R}^{4}$. 
After introducing the toroidal coordinates \cite{Frolov:2006pu}
\begin{align*}
  x&=\frac{\alpha\sin\hat{\theta}}{B}\cos\phi, \quad\quad y=\frac{\alpha\sin\hat{\theta}}{B}\sin\phi,\\
  z&=\frac{\alpha\sinh\eta}{B}\cos\psi, \quad w=\frac{\alpha\sinh\eta}{B}\sin\psi,
\end{align*}
on $\mathbb{R}^{4}$, where 
\[
B=\cosh\eta-\cos\hat\theta,
\qquad
 \alpha=\sqrt{L^2-r_H^2},
\]
the 4D Euclidean line element 
\[
{\mathrm d} s^{2}={\mathrm d}x^{2}+{\mathrm d}y^{2}+{\mathrm d}z^{2}
+{\mathrm d}w^{2}
\]
becomes
\begin{align}
  \mathrm{d} {s}^2 =\frac{\alpha^2}{B^2}\left[\mathrm{d} \eta^2
 +\sinh^2\eta\mathrm{d} \psi^2+\mathrm{d} \hat{\theta}^2+\sin^2\hat{\theta}\mathrm{d} \phi^2\right].
 \end{align}
The 3D hyper surface (\ref{hs}) corresponds to constant $\eta=\eta_0$, with
\begin{align*}
\cosh\eta_0=\frac{L}{r_H}.
\end{align*}
The line element on this 3-surface  is
\begin{align*}
  \mathrm{d} \tilde{s}^2 =\frac{L^2-r_H^2}{(L-r_H\cos\hat{\theta})^2}\left[(L^2-r_H^2)\mathrm{d} \psi^2 +r_H^2\mathrm{d} \hat{\theta}^2+r_H^2\sin^2\hat{\theta}\mathrm{d} \phi^2\right].
\end{align*}
After taking the coordinate transformation 
\begin{align*}
  \sin\theta=\frac{\sqrt{L^2-r_H^2}\sin\hat{\theta}}{L-r_H\cos\hat{\theta}},
\end{align*}
the above line element becomes (\ref{dst}). 

We can make the correspondence of the line element (\ref{dst}) with the 3D hyper surface (\ref{hs}) even more direct. To do this, we simply parametrize the 3-surface as
\begin{align*}
  x&=r_H\sin\theta\cos\psi,\\
  y&=r_H\sin\theta\sin\psi,\\
  z&=(L+r_H\cos\theta)\sin\phi,\\
  w&=(L+r_H\cos\theta)\cos\phi.
\end{align*}
In fact, the hypersurface equation (\ref{hs}) 
describes an $S^2$ fibration over $S^1$, with the circle $S^1$ parametrized by the
angle $\phi$. Therefore, the horizon 
surface is nothing but a conformally distorted $S^2$ bundle over $S^1$. Note that the 
hypersurface (\ref{hs}) is not of constant scalar curvature.

\subsection{Conformal factor}

To understand how seriously the conformal factor distorts the geometry of 
$S^2$ fibration over $S^1$, we need to study the behavior of the conformal factor. 
For any fixed $r=r_0$ (which is the case when we study the horizon geometry), the 
square root of the conformal factor, $(1+\alpha r_{0}\cos\theta)^{-1}$, sweeps an
ellipse if $\alpha r_0<1$, a parabola if $\alpha r_0=0$ or
a pair of hyperbola if $\alpha r_0>1$. The value of  $\alpha r_0$ corresponds to the
eccentricity of these conics. 

Since we shall be mainly interested in compact surfaces, we assume $\alpha r_0<1$. 
Under this condition, if the part of the horizon surface besides the conformal factor 
were a round sphere, then the effect of the conformal factor will simply be squashing the 
sphere into an ellipsoid \cite{Zhao:2011p693}. 
In the present case, however, the part of the horizon surface besides the 
conformal factor is an $S^2$ bundle over $S^1$, so the full horizon surface is 
a conformally squashed $S^2$ bundle over $S^1$. 

\subsection{Causal structure}

To make thorough understanding of the structure of the metric (\ref{5dm}), we now 
study its causal structure. First, we introduce the Eddington-Finkelstein coordinates,
\begin{align}
  u=t-r^*, \quad v=t+r^*,
\end{align}
where the tortoise coordinate $r^*$ is defined as
\begin{align}
  r^*=\int \bigg(1-\frac{\alpha r^2}{L}\bigg)^{-1} \mathrm{d} r
  =\frac{r_H}{2}\log  \left|\frac{r+r_H}{r-r_H}\right|,
\end{align}
and both $u$ and $v$ belong to the range $(-\infty, \infty) $. In this coordinate
system, the metric (\ref{5dm}) becomes
\begin{align}
  \mathrm{d} s^2 =\rho^2\left[-\bigg(1-\frac{\alpha r^2}{L}\bigg) \mathrm{d} u \mathrm{d} v
  + \mathrm{d}\Omega^2_3 \right],
\label{metric3}
\end{align}
where 
\begin{align}
\rho=(1+\alpha r\cos\theta)^{-1} \label{rho}
\end{align} 
and 
\begin{align*}
  \mathrm{d}\Omega^2_3=r^2\mathrm{d} \theta^2+r^2\sin^2\theta\mathrm{d} \psi^2+(L+r\cos\theta)^2\mathrm{d} \phi^2
\end{align*}
is 
the line element on the angular surface, whose geometry is very similar to that of the 
line element (\ref{dst}) which represents an $S^2$ bundle over $S^1$ as mentioned in Subsection \ref{hg}. 
The Kruskal coordinates are introduced as
\begin{align*}
\tilde{u}=\pm\exp\left(\frac{u}{r_H}\right),\quad\tilde{v}=\pm\exp\left(-\frac{v}{r_H}\right),
\end{align*}
where the signs of $\tilde{u}$ and $\tilde{v}$ coincide if $r<r_H$,
and are opposite to each other if $r\geq r_H$.
So there are totally 4 different combinations,
each of which corresponds to a causal patch in the conformal
diagrams to be drawn below. In each cases, one finds that
\begin{align}
  \tilde{u}\tilde{v}=-\frac{r-r_H}{r+r_H},
\end{align}
and eq.(\ref{metric3}) becomes
\begin{align}
    \mathrm{d} s^2 = \rho^2
    \left[-(r+r_H)^2\mathrm{d} \tilde{u} \mathrm{d}
    \tilde{v} + \mathrm{d}\Omega^2_3 \right],
\end{align}
where $r$ and $\rho$ are to be regarded as functions of
$\tilde{u}$ and $\tilde{v}$,
\begin{align*}
  r&=r_H\frac{1-\tilde{u}\tilde{v}}{1+\tilde{u}\tilde{v}},\\
  \rho&=\frac{1+\tilde{u}\tilde{v}}{(1+\tilde{u}\tilde{v}) +\alpha r_H(1-\tilde{u}\tilde{v})\cos\theta }.
\end{align*}
Finally, the Carter-Penrose coordinates can be introduced by the
usual arctangent mappings of $\tilde{u}$ and $\tilde{v}$
\begin{align*}
  U&=\arctan{\tilde{u}},\quad  V=\arctan{\tilde{v}},\\
  T&\equiv U+V, \quad  R\equiv U-V,
\end{align*}
in terms of which the metric becomes
\begin{align}
 \mathrm{d} s^2 = \frac{\rho^2r_H^2}{\cos^2 R}\left[- \mathrm{d} T^2 + \mathrm{d} R^2
 +\frac{1}{r_H^2}\cos^2 R ~\mathrm{d}\Omega^2_3 \right].
\end{align}

The values of the product
$\tilde{u}\tilde{v}$ at $r=0$ ($\rho=1$), $r=r_H$, $r=\infty$ ($\rho=0$)  and $\rho=\pm\infty$ are respectively
\begin{align*}
& \lim_{r \rightarrow 0} \tilde{u}\tilde{v}=1,\quad \Rightarrow U+V=\pm\frac{\pi}{2}\\
& \lim_{r \rightarrow r_H} \tilde{u}\tilde{v}=0,\quad \Rightarrow UV=0\\
& \lim_{r \rightarrow \infty} \tilde{u}\tilde{v}=-1,\quad \Rightarrow U-V=\pm\frac{\pi}{2}\\
& \lim_{\rho \rightarrow  \pm\infty} \ \tilde{u} \tilde{v}=-\frac{1+\alpha r_H\cos\theta}{1-\alpha r_H\cos\theta}.
  \end{align*}
The curves corresponding to $r=0$ , $r=r_H$ and $r=\infty$ can be easily depicted on the $(T,R)$ plane and they form various boundaries in the Carter-Penrose diagrams.
However, from (\ref{rho}), it follows that $\rho=\pm\infty$ will never be reachable if 
$\cos\theta\ge 0$. So, only for $\cos\theta<0$, the $\rho=\pm\infty$ curves on the
$(T,R)$ plane can possibly play as boundaries in the Carter-Penrose diagram. When this the case, these curves will separate our spacetime into two patches: the (+) patch with $\rho(r,\theta)>0$ and the $(-)$ patch with $\rho(r,\theta)<0$. Below we shall consider 3 different sub cases in detail.

\begin{figure}
\begin{center}
\includegraphics[width=1.00\textwidth]{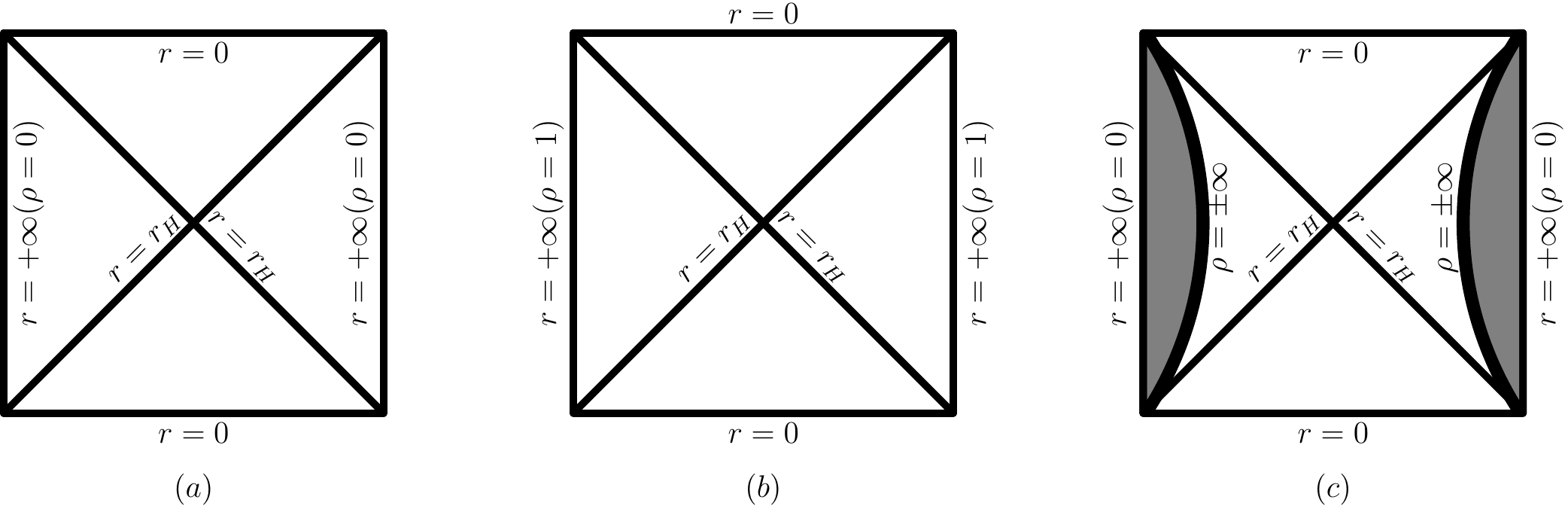}
\begin{minipage}{10.5cm}
\caption{Carter-Penrose diagrams: the shaded areas correspond to the $(-)$ patch, while the unshaded areas correspond to the (+) patch.} \label{Fig1}
\end{minipage}
\end{center}
\end{figure}

\begin{itemize}
\item $\cos\theta>0$ ($\theta\in[0, \pi/2)$). In this case, $\rho\in[0, 1)$ and there is no conformal infinity in the causal diagram. The Carter-Penrose diagram is consisted purely of the first 3 kinds of boundaries and is depicted in Fig.\ref{Fig1} (a).

\item $\cos\theta=0$ ($\theta=\pi/2$). In this case, $\rho$ takes the fixed value $1$, so is also no conformal infinities in the causal diagram. The Carter-Penrose diagram is depicted in Fig.\ref{Fig1} (b).

\item $\cos\theta<0$ ($\theta\in(\pi/2, \pi]$). In this case, $\rho\in(-\infty, 0]\cup(1, \infty)$. Moreover, the $\rho=+\infty$ and the $\rho=-\infty$ curves coincide on the $(T,R)$ plane, which separate the spacetime into (+) and $(-)$ patches.  It can be easily inferred from above that $-1<\lim_{\rho\rightarrow\pm\infty} \ \tilde{u} \tilde{v}<0$ and its value is varying with $\theta$. The corresponding Carter-Penrose diagram is depicted in Fig.\ref{Fig1} (c), in which the shaded areas correspond to the $(-)$ patch, while the unshaded area corresponds to the (+) patch. The variable $\rho$ is discontinuous in this case: it takes values in $(0,\infty)$ in the (+) patch and in $(-\infty,0]$ in the $(-)$ patch.
\end{itemize}

\section{Limiting cases}

The metric depends on two parameters $\alpha$ and $L$, which can be chosen to take various limits.

The first limit we can consider is taking $\alpha\rightarrow 0$ while keeping $L$ finite.
In this limit, both the conformal factor and the function $f(r)$ goes to unity and the 
metric becomes  
\begin{align*}
  \mathrm{d} s^2=-\mathrm{d} t^2+\mathrm{d} r^2+r^2 \mathrm{d} \theta^2
  +r^2\sin^2\theta \mathrm{d} \psi^2
  +(L+r\cos\theta)^2 \mathrm{d} \phi^2,
\end{align*}
which describes a Ricci flat spacetime. The spacial slices of the spacetime are 
nothing but a trivial fibration of $\mathbb{R}^{3}$ over $S^1$. The corresponding 
spacetime has vanishing proper accelerations for any static observer and there is no 
acceleration horizon. If, at this point, we require further $L\rightarrow 0$, then
the line element becomes that of the 5D Minkowski spacetime,
\begin{align*}
  \mathrm{d} s^2&=-\mathrm{d} t^2+\mathrm{d} r^2
  +r^2 \mathrm{d}\Omega_{3}^{2},
\end{align*}
where
\begin{align*}
  \mathrm{d}\Omega_{3}^{2}&=
  \mathrm{d}\theta^2
  +\sin^2\theta \mathrm{d} \psi^2
  +\cos^{2}\theta \mathrm{d} \phi^2
\end{align*}
is the metric of $S^{3}$ written in the standard Hopf coordinates. 
It is amazing that the presence of the acceleration parameter 
produces both the horizon and its nontrivial topology. 

The next limit we can consider is a double scaling limit $\alpha\rightarrow 0, L\rightarrow 0$ whilst $\alpha/L \rightarrow 1/l^2$ is kept finite. In this limit, the 
spacetime becomes the standard spherically symmetric de Sitter spacetime dS$_5$ with
radius $l$. Notice that we cannot take the limit $L\rightarrow 0$ while keeping 
$\alpha$ finite, because in that case the function $h(r)$ becomes undefined.

The third limit we can consider can be achieved by an overall rescaling of the metric.
Let us make a rescaling $t\rightarrow \sqrt{\frac{L}{\alpha}}\,\tilde t$, 
$r\rightarrow \sqrt{\frac{L}{\alpha}}\,\tilde r$, so that the resulting metric 
becomes
\begin{align}
  \mathrm{d} s^2&=\frac{A}{(1+\ell\,\tilde r\cos\theta)^2}
  \bigg\{-h(\tilde r) \mathrm{d} \tilde t^2+\frac{\mathrm{d}\tilde r^2}{h(\tilde r)}
  \nonumber\\
  &\qquad+\tilde r^2 \mathrm{d} \theta^2
  +\tilde r^2\sin^2\theta \,\mathrm{d} \psi^2
  +\left(\ell+\tilde r\cos\theta\right)^2 \mathrm{d} \phi^2 \bigg\},
  \label{lm2}
\end{align}
where 
\[
A=\frac{L}{\alpha},\qquad\ell=\sqrt{\alpha L},
\]
and
\begin{align*}
h(\tilde r)=1-\tilde r^2.
\end{align*}
Notice that the cosmological constant can be expressed in terms of $A$ and $\ell$
as
\[
\Lambda=6A^{-1}(1-\ell^2).
\]
So we have $0\le \ell<1$ for de Sitter case, $\ell=1$ for flat case and $\ell>1$ for AdS case. 

We can drop the overall constant factor $A$ in the line element,
which just results in a rescaling of the cosmological constant. The resulting metric will 
then depend only on a single parameter $\ell$. We can further take the limit
$\ell\rightarrow 0$, in which case the metric becomes that of the standard 5D de Sitter 
of unit radius. Note that this limit cannot be achieved if we had not made the rescaling
of the metric before.

\section{Euclideanization}

Euclideanization of Einstein vacua is an important subject of study from both physics and mathematical perspectives. The physics motivation for studying the Euclideanized
vacua is to study the vacuum transitions in Euclidean quantum gravity. The known 
examples of \cite{Page:1978p1271} and \cite{Hashimoto:2004p1291} were 
found for this purpose.  Mathematically, Euclideanized Einstein metrics often gives
explicit examples of smooth compact Riemannian manifolds, which is an important class 
of manifold in differential geometry. In the following, we shall see that there exist a 
Riemannian metric of constant scalar curvature in the conformal class of $S^4$ 
fibration over $S^1$. The corresponding metric is nothing but the Euclideanized version 
of our Einstein vacuum solution. 

Let us start from the line element (\ref{lm2}).
Making a Wick rotation $\tilde t\rightarrow i \chi$ and letting 
$\tilde r=\sin\sigma$, we get\footnote{By convention, we denote the Euclideanized 
line elements with a bar and the dimension of the Euclideanized space is specified by
a suffix.}
\begin{align}
  \mathrm{d} \bar s_5^2&=\frac{A}{(1+\ell\,\sin\sigma\cos\theta)^2}
  \bigg\{\mathrm{d}\sigma^2+\cos^2\sigma \,\mathrm{d}\chi^2
  \nonumber\\
  &\qquad+\sin^2\sigma \,(\mathrm{d} \theta^2
  +\sin^2\theta\, \mathrm{d} \psi^2)
  +\left(\ell+\sin\sigma\cos\theta\right)^2 \mathrm{d} \phi^2 \bigg\}.
  \label{euc}
\end{align}

In the absence of the conformal factor, this metric is just an $S^4$ fibration over $S^1$ 
which is not a constant curvature manifold\footnote{That the metric (\ref{euc}) 
without the 
conformal factor describes an $S^4$ fibration over $S^1$ can also be 
understood from an extrinsic geometric point of view, just like the way we understood
the horizon geometry in Section 2. We put this extrinsic geometric description 
in the appendix.}. The coordinate $(\sigma,\chi,\theta,\psi)$ parametrizes the $S^{4}$ 
fiber, which can be regarded as a generalization of the standard Hopf coordinates for 
$S^{3}$. The ranges of these coordinates are given in (\ref{range}) in the appendix.
The conformal factor squashes the round $S^4$ fiber and makes 
the Ricci scalar of the full space a constant. In other words, there is a constant scalar 
curvature manifold in the conformal class of $S^4$ fibration over $S^1$. The study 
of conformal class of a given Riemannian manifold is an important subject of study in 
differential geometry, because this subject is intimately related to the analysis of 
geometric flows.

Since the metric (\ref{euc}) represents a conformally distorted $S^4$ fibration over 
$S^1$, we shall refer to the corresponding geometry as a ``ring geometry'', with the ring
(i.e. the $S^1$ factor) parametrized by the angle $\phi$. Fibers of the ring geometry at 
a given $\phi$ is a conformal 4-sphere. The overall constant $A$ signifies the size of the 
ring surface, and the parameter $\ell$ represents the relative radius of the $S^{1}$ with 
respect to that of the $S^{4}$. We have mentioned previously that when 
$0<\ell<1$, the corresponding geometry is smooth, compact and inhomogeneous, with a 
positive constant scalar curvature. Perhaps it is the first known example of such metrics 
on $S^4$ fibration over $S^1$. The case $\ell=0$ corresponds to a degenerate 
case, i.e. a conformal $S^5$.


It is interesting to evaluate the volume of this compact space. Without loss of generality, 
we set $A=1$ in the following calculations. We have, by direct calculation, 
\begin{align*}
\sqrt{g}=
\frac{\cos\sigma\sin^2\sigma\sin\theta\left(\ell
+\sin\sigma\cos\theta\right)}{\left(1+\ell\sin\sigma
\cos\theta\right)^5}.
\end{align*}
Using the coordinate range (\ref{range}), it is easy to evaluate the 5-volume of the 
space. The result is
\[
V_5=\int d^5x \sqrt{g}=0.
\]
Since the space is of constant scalar curvature, the corresponding Einstein-Hilbert action
is proportional to the 5-volume, so its value is also zero. Compact 
Einstein metrics of zero volume are not rare. See \cite{Wang:1990p1463} for other 
examples.

\section{General dimensions}

So far we have been restricting ourselves in five dimensions. As mentioned earlier, 5D 
is not of any particular importance in the construction. There is a higher
dimensional cousin for the metric we studied in the previous sections.
The dimensional metric can be written as
\begin{align*}
  \mathrm{d} s_D^2&=\frac{1}{(1+\alpha r\cos\theta)^2}\bigg\{-h(r) \mathrm{d} t^2+\frac{\mathrm{d} r^2}{h(r)}\\
  &\qquad +r^2 \mathrm{d} \theta^2
  +r^2\sin^2\theta \,\mathrm{d} \Omega_{D-4}^2
  +(L+r\cos\theta)^2 \mathrm{d} \phi^2 \bigg\},
\end{align*}
where $h(r)$ is as before, and $\mathrm{d} \Omega_{D-4}^2$ is the line element of a round $(D-4)$-sphere. The associated cosmological constant is given by
\[
\Lambda_D=\frac{(D-1)(D-2)}{2}\frac{\alpha}{L}(1-\alpha L).
\]
The physical and mathematical properties of this metric is extremely similar to the 5D
case. For instance, the metric interpreted as Einstein vacuum represents an accelerating 
vacuum in which the acceleration horizon has the topology of $S^{D-3}$
fibration over $S^1$. The Euclideanized version of the metric represents a  
conformally distorted metric on the full space of an $S^{D-1}$ fibration over $S^1$ 
and this metric is only compact for small value of $\ell$, etc. 

The Euclideanized version of the higher dimensional metric is given as follows,\begin{align}
  \mathrm{d} \bar s_D^2&=\frac{A}{(1+\ell\,\sin\sigma\cos\theta)^2}
  \bigg\{\mathrm{d}\sigma^2+\cos^2\sigma \,\mathrm{d}\chi^2
  \nonumber\\
  &\qquad+\sin^2\sigma \,(\mathrm{d} \theta^2
  +\sin^2\theta\, \mathrm{d} \Omega_{D-4}^2)
  +\left(\ell+\sin\sigma\cos\theta\right)^2 \mathrm{d} \phi^2 \bigg\}.
  \label{euc2}
\end{align}
This metric describes a conformally squashed $S^{D-1}$ fibration over $S^{1}$. 
The cosmological constant can be written in terms of the parameters $A$ and $\ell$
as
\begin{align}
\Lambda_D=\frac{(D-1)(D-2)}{2}\left(\frac{1-\ell^2}{A}\right). \label{lD}
\end{align}

{\em \underline{Remarks}:}

The metric (\ref{euc2}) makes sense for all dimensions $D\ge 3$. We have already 
analyzed in detail the $D=5$ case. For $D=4$, we just remove the 
$\mathrm{d}\Omega_{D-4}^2$ term from the metric and recovers the metric given in
\cite{Astorino:2011p1101}. For $D=3$, we remove
the $\mathrm{d}\theta^2$ term altogether, and set $\theta=0$ everywhere else.

\section{Conclusion and discussion}

We presented a novel class of accelerating Einstein vacua with accelerating horizon bearing a nontrivial topology of $S^{D-3}$ bundle over $S^{1}$. Such solutions contain
two parameters, one corresponds to the acceleration, the other corresponds to the relative 
radius of the $S^{1}$ base with respect to the $S^{D-3}$ fiber. There are various 
limiting cases for these parameters. Among these, the zero acceleration limit corresponds 
to a Ricci flat vacuum with no horizon. The double scaling limit gives rise to the 
standard de Sitter vacua. Upon Euclideanization, the full space becomes a smooth 
compact inhomogeneous Riemannian manifold with a positive constant scalar curvature.  
Such Euclidean manifolds can be regarded as analogues of Page's gravitational instanton 
or its generalizations, but the topologies are now conformally distorted $S^{D-1}$ 
bundle over $S^{1}$, rather than sphere bundles over $S^{2}$.

Since for generic values of the parameters, the vacua we obtained possess  a positive cosmological constant and a ring-like acceleration horizon, we expect that such vacua 
should be the starting point to construct black rings with cosmological constant. 
Asymptotically flat black ring solutions in Einstein gravity were found over ten years ago,
but so far no black ring solutions with cosmological constant were found. 
One of the major obstacle for constructing such solutions is the un-matching 
topologies: black ring solutions have horizons with $S^{2}\times S^{1}$ topology, whilst 
the usual 5D de Sitter spacetime has only an accelerating horizon of $S^{3}$ topology.
The new vacua we found in this article do have the matching topology with black rings.
To actually construct black rings which asymptote to our vacuum solutions, 
some complicated mathematical constructions are yet to be carried out. Presumably the
Kerr-Schild method which led to the discoveries of rotating black holes in higher 
dimensions with cosmological constants \cite{Gibbons:2004p534,Gibbons:2004p533} 
is a good starting point. We shall continue our explorations in this direction.

\section*{Appendix: extrinsic geometric description of $S^{4}$ fibration 
over $S^{1}$}

Consider the following surface
\begin{align}
  x_{1}^2+x_{2}^2+x_{3}^2+x_{4}^2+(\sqrt{x_{5}^2+x_{6}^2}-\ell)^2=1 
  \label{s4xs1}
\end{align}
in the 6D Euclidean space with metric $\mathrm{d}s^{2}
=\sum_{i=1}^{6}dx_{i}^{2}$. 
It is straightforward to parametrize this surface as
\begin{align*}
  x_{1}&=\cos\sigma\sin\chi,\\
  x_{2}&=\cos\sigma\cos\chi,\\
  x_{3}&=\sin\sigma\sin\theta\cos\psi,\\
  x_{4}&=\sin\sigma\sin\theta\sin\psi,\\
  x_{5}&=(\ell+\sin\sigma\cos\theta)\sin\phi,\\
  x_{6}&=(\ell+\sin\sigma\cos\theta)\cos\phi,
\end{align*}
where the angular coordinates $(\sigma, \chi, \theta, \psi, \phi)$ must be chosen such 
that these parametrization equations cover the surface (\ref{s4xs1}) exactly once. The 
ranges for these angular coordinates are given as follows:
\begin{align}
\begin{matrix}
\sigma\in[0,\pi/2], &\theta\in [0,\pi],&\cr
\chi\in[0,2\pi],&\psi\in[0,2\pi],&\phi\in[0,2\pi].
\end{matrix}  \label{range}
\end{align}

Inserting the parametrization equations into the 6D Euclidean line element we get the
desired result 
\begin{align*}
\mathrm{d}s^{2}=\mathrm{d}\sigma^2+\cos^2\sigma \,\mathrm{d}\chi^2
  +\sin^2\sigma \,(\mathrm{d} \theta^2
  +\sin^2\theta\, \mathrm{d} \psi^2)
  +\left(\ell+\sin\sigma\cos\theta\right)^2 \mathrm{d} \phi^2,
\end{align*}
which is identical to the line element (\ref{euc}) without the conformal 
factor. The Ricci scalar associated with the above metric reads
\begin{align*}
R=\frac{4(3\ell+5\sin\sigma\cos\theta)}{\ell+\sin\sigma\cos\theta},
\end{align*}
which is not a constant. The surface (\ref{s4xs1}) clearly describes an $S^{4}$ fibration 
over $S^{1}$.

\section*{Acknowledgment} 

This work is supported in part by the National Natural Science Foundation of 
China (NSFC) through grant No.10875059.


\begin{thebibliography}{10}

\bibitem{Horne:1991gn}
  J.~H.~Horne, G.~T.~Horowitz,
  ``Exact black string solutions in three-dimensions,''
  Nucl.\ Phys.\  {\bf B368 } (1992)  444-462.
  [arXiv:\href{http://www.arXiv.org/abs/hep-th/9108001}{{\tt
  hep-th/9108001}}].

\bibitem{Anderson:1995hu}
  W.~G.~Anderson, N.~Kaloper,
  ``On some new black string solutions in three-dimensions,''
  Phys.\ Rev.\  {\bf D52 } (1995)  4440-4454.
  [arXiv:\href{http://www.arXiv.org/abs/hep-th/9503175}{{\tt
  hep-th/9503175}}].
  
\bibitem{Mahapatra:1993gx}
  S.~Mahapatra,
  ``On the rotating charged black string solution,''
  Phys.\ Rev.\  {\bf D50 } (1994)  947-951.
  [arXiv:\href{http://www.arXiv.org/abs/hep-th/9301125}{{\tt
  hep-th/9301125}}].

\bibitem{Emparan:2001p607}
R.~Emparan and H.~S. Reall, ``Generalized Weyl Solutions,'' Phys.Rev. D65 (2002) 084025 [arXiv:\href{http://www.arXiv.org/abs/hep-th/0110258}{{\tt
  hep-th/0110258}}].

\bibitem{Emparan:2001p762}
R.~Emparan and H.~S. Reall, ``A rotating black ring in five dimensions,'' Phys.Rev.Lett.88:101101,2002 [arXiv:\href{http://www.arXiv.org/abs/hep-th/0110260}{{\tt hep-th/0110260}}].

\bibitem{Elvang:2004p3129}
H.~Elvang, R.~Emparan, D.~Mateos, and H.~S. Reall, ``A supersymmetric black
  ring,'' Phys.Rev.Lett.93:211302,2004 [arXiv:\href{http://www.arXiv.org/abs/hep-th/0407065}{{\tt hep-th/0407065}}].

\bibitem{Emparan:2006p2455}
R.~Emparan and H.~S. Reall, ``Black Rings,'' Class.Quant.Grav.23:R169,2006 [arXiv:\href{http://www.arXiv.org/abs/hep-th/0608012}{{\tt
  hep-th/0608012}}].

\bibitem{Zhao:2011p693}
L.~Zhao, ``Note on a class of anisotropic Einstein metrics,'' [arXiv:\href{http://www.arXiv.org/abs/1106.5027}{{\tt
  1106.5027}}]. 
  
\bibitem{Zhao:2011p1230}
L.~Zhao and K.~Meng, ``Gauss-Bonnet as effective cosmological constant,'' Commun. Theor. Phys. 57 (2012) 607–610 [arXiv:\href{http://www.arXiv.org/abs/1109.6748}{{\tt 1109.6748}}]. 

\bibitem{Xu:2011p1323}
W.~Xu, K.~Meng, and L.~Zhao, ``Accelerating vacua in Gauss-Bonnet gravity,''
Commun. Theor. Phys. (2012) to appear
  [arXiv:\href{http://www.arXiv.org/abs/1110.5769}{{\tt 1110.5769}}]. 

\bibitem{Mann:1996p968}
R.~B. Mann, ``Pair Production of Topological anti de Sitter Black Holes,'' 
Class.Quant.Grav.14:L109-L114,1997 [arXiv:\href{http://www.arXiv.org/abs/gr-qc/9607071}{{\tt gr-qc/9607071}}].

\bibitem{Vanzo:1997p1331}
L.~Vanzo, ``Black holes with unusual topology,'' Phys.Rev. D56 (1997) 6475-6483
 [arXiv:\href{http://www.arXiv.org/abs/gr-qc/9705004}{{\tt gr-qc/9705004}}].

\bibitem{Birmingham:1998p1113}
D.~Birmingham, ``Topological Black Holes in Anti-de Sitter Space,'' 
 Class.Quant.Grav. 16 (1999) 1197-1205 [arXiv:\href{http://www.arXiv.org/abs/hep-th/9808032}{{\tt hep-th/9808032}}].

\bibitem{Page:1978p1271}
D.~N. Page, ``A compact rotating gravitational instanton,'' Phys.
  Lett. B {\bf 79} (1978), no.~3, 235--238.

\bibitem{Hashimoto:2004p1291}
Y.~Hashimoto, M.~Sakaguchi, and Y.~Yasui, ``New Infinite Series of Einstein
  Metrics on Sphere Bundles from AdS Black Holes,'' 
  Commun.Math.Phys. 257 (2005) 273-285 [arXiv:\href{http://www.arXiv.org/abs/hep-th/0402199}{{\tt
  hep-th/0402199}}].
  
\bibitem{Astorino:2011p1101}
M.~Astorino, ``Accelerating black hole in 2+1 dimensions and 3+1 black
  (st)ring,'' 	JHEP 1101:114,2011 [arXiv:
  \href{http://www.arXiv.org/abs/1101.2616}{{\tt 1101.2616}}].


\bibitem{Frolov:2006pu}
  V.~P.~Frolov and R.~Goswami,
  ``Surface Geometry of 5D Black Holes and Black Rings,''
  Phys.\ Rev.\ D {\bf 75} (2007) 124001  [gr-qc/0612033].


\bibitem{Wang:1990p1463}
M.~Wang and W.~Ziller, ``Einstein metrics on principal torus bundles,'' 
  J. of Diff. Geom. {\bf 31} (1990), no.~1, 215--248.

\bibitem{Gibbons:2004p534}
G.~W. Gibbons, D.~N. Page, and C.~N. Pope, ``Rotating Black Holes in Higher
  Dimensions with a Cosmological Constant,'' Phys.Rev.Lett.93:171102,2004 
  [arXiv:\href{http://www.arXiv.org/abs/hep-th/0409155}{{\tt
  hep-th/0409155}}].

\bibitem{Gibbons:2004p533}
G.~W. Gibbons, D.~N. Page, and C.~N. Pope, ``The General Kerr-de Sitter Metrics
  in All Dimensions,'' J.Geom.Phys.53:49-73,2005 [arXiv:\href{http://www.arXiv.org/abs/hep-th/0404008}{{\tt hep-th/0404008}}].
  

\end{thebibliography}

\providecommand{\href}[2]{#2}\begingroup\raggedright\endgroup

\end{document}